\documentclass[twocolumn,epsf,graphics,psfig,superscriptaddress]{revtex4}

\usepackage{graphicx,graphics} 
\usepackage[dvips,bookmarks=false]{hyperref}
\usepackage{graphicx}
\usepackage{dcolumn}
\usepackage{eucal}
\usepackage[dvips]{epsfig}

\begin{document}
\title{Spin currents and magnetoresistance of graphene-based magnetic junctions}
\author{Alireza Saffarzadeh} \email{a-saffar@tehran.pnu.ac.ir}
\affiliation{Department of Physics, Payame Noor University,
Nejatollahi Street, 159995-7613 Tehran, Iran}
\affiliation{Computational Physical Sciences Laboratory,
Department of Nano-Science, Institute for Research in Fundamental
Sciences (IPM), P.O. Box 19395-5531, Tehran, Iran}
\author{Mahdi Ghorbani Asl}
\affiliation{Department of Physics, Islamic Azad University, North
Tehran Branch, Darband Street, Tehran, Iran }
\date{\today}

\begin{abstract}
Using the tight-binding approximation and the nonequilibrium
Green's function approach, we investigate the coherent
spin-dependent transport in planar magnetic junctions consisting
of two ferromagnetic (FM) electrodes separated by a graphene flake
(GF) with zigzag or armchair interfaces. It is found that the
electron conduction strongly depends on the geometry of contact
between the GF and the FM electrodes. In the case of zigzag
interfaces, the junction demonstrates a spin-valve effect with
high magnetoresistance (MR) ratios and shows negative differential
resistance features for a single spin channel at positive gate
voltage. In the case of armchair interfaces, the current-voltage
characteristics behave linearly at low bias voltages and hence,
both spin channels are in on state with low MR ratios.
\end{abstract}
\maketitle

\section{Introduction}
Spin-dependent transport through magnetic nanojunctions is
currently receiving increased attention owing to the expectation
that nanoscale size constraints may maximize the potentially
useful magnetoresistance effects. Among various types of nanoscale
devices, carbon-based nanostructures are attractive for
spin-polarized electronics \cite{Wolf,Moodera} because the
relatively weakness of spin-orbit and hyperfine interactions
should lead to long spin coherence times \cite{Sanvito}. In
particular, the spin-dependent transport studies on carbon
nanotubes coupled to ferromagnetic Co electrodes have shown that
nanotubes exhibit a considerable tunnel MR effect
\cite{Tsukagoshi,Chakraborty}. A C$_{60}$ molecule is also an
interesting nanocarbon unit, and recently the spin-dependent
electron transport for the granular film consisting of Co
nanoparticles embedded in a Co-C$_{60}$ compound matrix has been
studied, and the maximum MR ratio of about 30\% at low bias
voltages was reported \cite{Zare,Sakai,Miwa}.

Another related carbon-based structure is graphene which
demonstrates many useful electronic properties: high carrier
mobility, long spin relaxation times and lengths, extreme
flexibility and stability, and metallic properties controllable by
a gate electrode \cite{Novoselov,Geim}. Moreover, due to the flat
structure, graphene seems to be easier to manipulate than carbon
nanotubes. Furthermore, interesting properties appear if the
graphene is patterned into ribbon like geometry called the
graphene nanoribbon (GNR). For instance, they can be either
semiconducting with a size dependent gap or metallic \cite{Han}.
Transport properties of GNRs are expected to depend strongly on
whether they have an armchair or zigzag edge \cite{Nakada}. In
GNRs with zigzag edges, transport is dominated by edge states
which have been observed in scanning tunneling microscopy
\cite{Kobayashi}. These states are expected to be spin-polarized
and make zigzag GNRs attractive for spintronics \cite{Prinz,Wolf}.

Based on first-principles calculations, Son \textit{et al.}
\cite{Son} predicted that the zigzag GNRs become half-metallic
when an external transverse electric field is applied. Using such
calculations, Kim and Kim \cite{Kim} showed that zigzag GNRs
connected to two ferromagnetic (FM) electrodes exhibit very large
values of magnetoresistance. Using the tight-binding
approximation, Bery and Fertig \cite{Bery} studied the
magnetoresistance of GNR-based spin valves in the infinite width
limit. They reported a feeble magnetoresistance due to the weak
dependence of the graphene conductivity on the electronic
parameters of the FM leads. The other spin-dependent properties of
graphene such as spin field effect transistor \cite{Semenov}, spin
Hall effects \cite{Kane,Sinitsyn}, and magnetic ordering in
graphene \cite{Okada} have also been predicted.

Experimentally, based on the nonlocal magnetoresistance
measurements, spin injection into a graphene thin film connected
to FM electrodes has been successfully demonstrated
\cite{Tombros,Cho,Ohishi} and the possibility of spin transport
and spin precession over micrometer-scale distances at room
temperature reported \cite{Tombros}. Hill \textit{et al.}
\cite{Hill} have observed large magnetoresistances of several
hundred ohms in a spin-valve device where a 200 nm wide graphene
wire is contacted by two soft magnetic NiFe electrodes. Also, Wang
\textit{et al.} \cite{Wang} have investigated magnetotransport
properties of spin valves consisting of graphite flakes connected
to FM electrodes, and the magnetoresistance values up to 12\% at 7
K were observed when an ultrathin MgO tunnel barrier was inserted
at the FM/graphite interface. These experiments clearly
demonstrate the possibility of spintronics applications for
graphene.

In this paper, we present a theory to investigate the
spin-dependent properties of planar magnetic junctions consisting
of a graphene flake (GF) connected by two FM electrodes in the
presence of bias and gate voltages. A GF is a finite rectangular
GNR with finite number of carbon atoms and, therefore, it can be
coupled to the two square-lattice FM electrodes through its
armchair or zigzag edges \cite{Robinson,Blanter}. Such
configurations have been shown in Fig. 1. The present approach
indicates that the spin-polarized transport in such magnetic
junctions is mainly controlled by the molecular field of the FM
electrodes, the geometry of FM/GF interface, bias and gate
voltages, and the strength of FM/GF coupling.
\begin{figure}
\centerline{\includegraphics[width=1.0\linewidth]{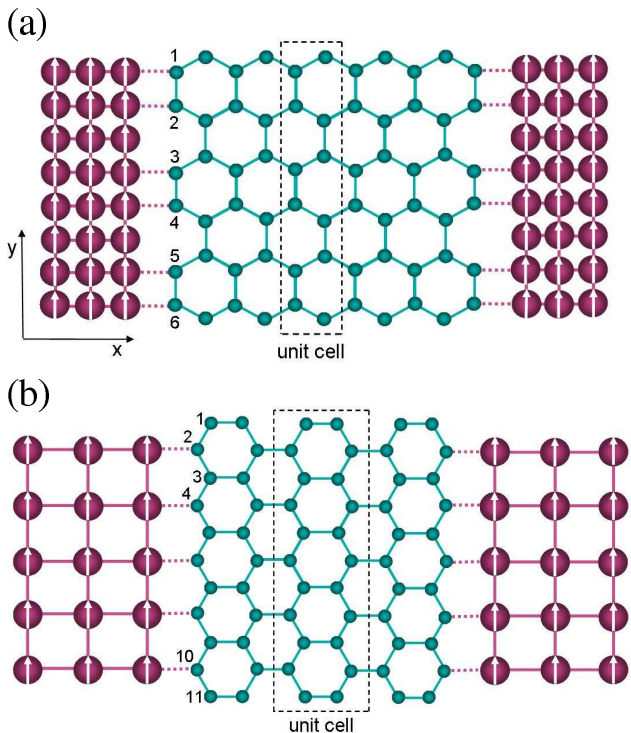}}
\caption{Schematic view of two FM/GF/FM magnetic junctions with
(a) armchair and (b) zigzag interfaces, at zero gate and bias
voltages. In both cases, the magnetic alignments of the electrodes
are shown in the parallel configuration. For the antiparallel
configuration, the magnetic alignment of the right (drain)
electrode should be reversed.}
\end{figure}
The paper is organized as follows. In Sec. II, we present the
model and formalism for spin transport through FM/GF/FM magnetic
junctions where the interfaces between FM leads and GF can be
armchair or zigzag. Numerical results and discussions for
transmission coefficients, spin currents and magnetoresistance are
presented in Sec. III. We conclude our findings in Sec. IV.

\section{Model and formalism}
We consider a system which consists of a rectangular GF with
armchair and zigzag edges, and two square-lattice FM electrodes.
The GF can be coupled to the magnetic electrodes in such a way
that their interface is armchair or zigzag. The two models for
such structures are shown schematically in Fig. 1. Since the
electron conduction is mainly determined by the graphitic region,
the electronic structure of this part should be resolved in
detail. Hence, we decompose the total Hamiltonian of the system as
\begin{equation}
\hat{H}=\hat{H}_{L}+\hat{H}_{C}+\hat{H}_{R}+\hat{H}_{T}\  ,
\end{equation}
where $\hat{H}_L$ and $\hat{H}_R$ are the Hamiltonians of the left
($L$) and right ($R$) electrodes, $\hat{H}_C$ describes the
Hamiltonian of the graphitic region (GF), and $\hat{H}_T$ is the
coupling of the GF to the electrodes. All terms of the total
Hamiltonian are described within the single-band tight-binding
approximation and can be written as
\begin{equation}
\hat{H}_{\alpha}=\sum_{<i_\alpha,
j_\alpha>,\sigma}[(\epsilon_{\alpha}-\mathbf{\sigma}\cdot
\mathbf{m}_{\alpha})\delta_{i_\alpha,j_\alpha}-t_{i_\alpha,j_\alpha})]\,\hat{c}_{i_\alpha,\sigma}^\dag\hat{c}_{j_\alpha,\sigma}\
,
\end{equation}
\begin{equation}
\hat{H}_{C}=\sum_{<i_C,j_C>,\sigma}(\epsilon_{_C}\delta_{i_C,j_C}-t_{i_C,j_C})\,\hat{d}_{i_C,\sigma}^\dag\hat{d}_{j_C,\sigma}\
,
\end{equation}
\begin{equation} \hat{H}_{T}=-\sum_{\alpha=\{L,R\}}\sum_{i_\alpha,
j_C,\sigma}t_{i_\alpha,j_C}(\hat{c}_{i_\alpha,\sigma}^\dag\hat{d}_{j_C,\sigma}+\textrm{H.c.})\
\ ,
\end{equation}
where $\hat{c}_{i_\alpha,\sigma}^\dag$
($\hat{c}_{i_\alpha,\sigma}$) and $\hat{d}_{i_C,\sigma}^\dag$
($\hat{d}_{i_C,\sigma}$) create (annihilate) an electron with spin
$\sigma$ on site $i$ in the electrodes and the graphitic region,
$t_{i_\alpha,j_\alpha}=t_L$, $t_{i_C,j_C}=t_C$, and
$t_{i_\alpha,j_C}=t_{LC}$ if $i$ and $j$ are nearest neighbors and
zero otherwise. $\epsilon_{\alpha}$, the on-site energy of the
electrodes, is equal to $2t_L$ which acts as a shift in energy,
and $\epsilon_{_C}$, the on-site energy of the GF, is zero except
in the presence of gate voltage $V_G$, that shifts the energy
levels of the GF. Here, $-\mathbf{\sigma}\cdot
\mathbf{m}_{\alpha}$ is the internal exchange energy with
$\mathbf{m}_{\alpha}$ denoting the local magnetization at site
$i_\alpha$, and $\mathbf{\sigma}$ being the conventional Pauli
spin operator.

In this study we set $t_L=t_C=t$ and assume that the transport is
ballistic \cite{Datta}; hence, we set $t_{LC}=0.8\,t$, because the
value of $t_{LC}$ should not be smaller than the order of $t$.
Such a ballistic approximation is valid when the mean free path of
the carriers is greater than the sample dimensions. High mobility
of charge carriers in graphene, even at room temperature, implies
that carriers can move long distances without scattering
\cite{Novoselov}. Hence, we expect that the ballistic
approximation to be appropriate in studying transport in GFs. In
addition, we assume that the spin direction of the electron is
conserved in the tunneling process through the graphitic region.
Therefore, there is no spin-flip scattering and the spin-dependent
transport can be decoupled into two spin currents: one for spin-up
and the other for spin-down. This assumption is well-justified
since the spin diffusion length in organics is about 4 nm
\cite{Sanvito} and especially in carbon nanotubes is at least 130
nm \cite{Tsukagoshi} and in graphene is 1.5 $\mu$m \cite{Jozsa},
which are greater than the length scale of our device. Since the
total Hamiltonian does not contain inelastic scatterings, the spin
currents for a constant bias voltage, $V_a$, are calculated by the
Landauer-B\"{u}ttiker formula based on the nonequilibrium Green's
function method \cite{Datta}:
\begin{equation}\label{I}
I_\sigma(V_a)=\frac{e}{h}\int_{-\infty}^{\infty}
T_\sigma(\epsilon,V_a)[f(\epsilon-\mu_L)-f(\epsilon-\mu_R)]d\epsilon
\ ,
\end{equation}
where $f$ is the Fermi-Dirac distribution function,
$\mu_{L,R}=E_F\pm \frac{1}{2}eV_a$ are the chemical potentials of
the electrodes, and
$T_\sigma(\epsilon,V_a)=\mathrm{Tr}[\hat{\Gamma}_{L,\sigma}
\hat{G}_\sigma \hat{\Gamma}_{R,\sigma}\hat{G}_\sigma^{\dagger}]$
is the spin-, energy- and voltage-dependent transmission function.
The spin-dependent Green's function of the graphitic region
coupled to the two FM electrodes (source and drain) in the
presence of the bias voltage is given as
\begin{equation}
\hat{G}_\sigma(\epsilon,V_a)=[\epsilon \hat{1}
-\hat{H}_C-\hat{\Sigma}_{L,\sigma}(\epsilon-eV_a/2)
-\hat{\Sigma}_{R,\sigma}(\epsilon+eV_a/2)]^{-1}\  ,
\end{equation}
where $\hat{\Sigma}_{L,\sigma}$ and $\hat{\Sigma}_{R,\sigma}$
describe the self-energy matrices which contain the information of
the electronic structure of the FM electrodes and their coupling
to the GF. These can be expressed as $
\hat{\Sigma}_{\alpha,\sigma}(\epsilon)=
\hat{\tau}_{C,\alpha}\hat{g}_{\alpha,\sigma}(\epsilon)\hat{\tau}_{\alpha,C}$
where $\hat{\tau}$ is the hopping matrix that couples the graphene
to the leads and depends on the geometry of the FM/GF interface.
$\hat{g}_{\alpha,\sigma}$ are the surface Green's functions of the
uncoupled leads i.e., the left and right semi-infinite magnetic
electrodes, and their matrix elements are given by
\begin{equation}\label{g}
g_{\alpha,\sigma}(i,j;\epsilon)=\sum_{k_x,l_y}\frac{\psi_{k_x,l_y}(x_i,y_i)\psi^*_{k_x,l_y}(x_j,y_j)}
{\epsilon+i\delta-\epsilon_\alpha+\mathbf{\sigma}\cdot
\mathbf{m}_{\alpha}+\varepsilon({k_x,l_y})}\  ,
\end{equation}
where $\delta$ is a positive infinitesimal,
$\varepsilon({k_x,l_y})=2t[\cos(k_xa)+\cos(\frac{l_y\pi}{N_y+1})]$,
and
\begin{equation}
\psi_{k_x,l_y}(x_i,y_i)=\frac{2}{\sqrt{N_x(N_y+1)}}
\sin(k_xx_i)\sin(\frac{l_yy_i\pi}{N_y+1})\ .
\end{equation}
Here, $l_{y}$ ($=1,...,N_{y}$) is an integer,
$k_x\in[-\frac{\pi}{a},\frac{\pi}{a}]$, and $N_\beta$ with $\beta
=x, y$ is the number of lattice sites in the $\beta$ direction.
Using $\hat{\Sigma}_{\alpha,\sigma}$, the coupling matrices
$\hat{\Gamma}_{\alpha,\sigma}$, also known as the broadening
functions, can be expressed as
$\hat{\Gamma}_{\alpha,\sigma}=-2\mathrm{Im}(\hat{\Sigma}_{\alpha,\sigma})$.

When the GF is brought close to an electrode, the bonding between
them will depend on the arrangement and the nature of the atoms at
the FM/GF interface. In Fig. 1(a), the graphitic region consisting
of 5.5 unit cells is matched to the FM leads along its armchair
edges. In this case, the lattice constant of the leads, $a_L$, is
equal to the lattice constant of the GF, $a_C$. On the other hand,
if two FM electrodes are coupled to the graphene along its zigzag
edges, we obtain Fig. 1(b) where $a_L=\sqrt{3}\,a_C$ and the GF
consists of 3 unit cells. In both junctions, there is the same
number of carbon atoms at the graphitic part, while the number of
contact points at the FM/GF interfaces has decreased from six (in
Fig. 1(a)) to five (in Fig. 1(b)). Therefore, one can expect
different spin transport through the GF, which will be discussed
in the next section. We should note that, the core of the problem
lies in the calculation of the spin-dependent self-energies
$\hat{\Sigma}_{L,\sigma}$ and $\hat{\Sigma}_{R,\sigma}$. In this
regard, for a GF with $N$ carbon atoms and in the case of contact
through $n_a$ ($n_z$) carbon atoms at the armchair (zigzag)
interfaces, only $n_a^2$ ($n_z^2$) elements of the self-energy
matrices will be non-zero.

\section{Results and discussion}
We now use the method described above to study the coherent
spin-dependent transport and magnetoresistance effect of FM/GF/FM
junctions with armchair and zigzag interfaces. We have done the
numerical calculations for the case that the direction of
magnetization in the left FM electrode is fixed in the +$y$
direction, while the magnetization in the right electrode is free
to be flipped into either the +$y$ or -$y$ direction. We choose a
GF with $N$=66 carbon atoms and set $t_L=t_C$=1 eV,
$|\mathbf{m}_{\alpha}|$=0.75 eV, $E_F$=0.0 eV, and $T$=300 K in
the calculations.

In Figs. 2 and 3 we depict the spin-dependent transmission
coefficients for the junctions with zigzag ($n_z$=5) and armchair
($n_a$=6) interfaces, respectively. Only the energy window around
the Fermi level has been shown. In these figures, the transmission
characteristics clearly demonstrate the dependence of electron
conduction on the geometry of contact between the FM electrodes
and the GF. In the case of parallel alignment of the junction with
zigzag interfaces, there are non-zero transmissions only for
spin-up electrons (on state) and the spin-down ones are in off
state. In the case of antiparallel alignment, both spin channels
are in off state. This is the so-called spin-valve effect or
magnetoresistive effect and indicates that, the junction with
zigzag interfaces can work as a switch, or a bit with on and off
states. The spin-valve devices are promising candidates for
systems that may transform spin information into electrical
signals.

From the spin-dependent currents of the junction with zigzag
interfaces it is clear that at low applied voltages ($V_a\leq 0.5$
V) only the spin-up current in the parallel alignment is in on
state (see Figs. 4(a)-(c)). The current-voltage characteristics of
such electrons are linear at low voltages and display different
behaviors at higher voltages, depending on the value of gate
voltage. One interesting feature of the $I$-$V$ characteristics is
that the spin-up current for $V_G$=0.4 V shows a negative
differential resistance region which is visible as a decrease in
the current upon increasing $V_a$ \cite{Dragoman}. This behavior
is due to a gradual disappearance of a resonant level within the
voltage window in that region. In addition, the appearance of zero
current and the necessity of threshold voltage to generate finite
current flow through the junction in some of the $I$-$V$ curves,
arising from the energy mismatch between the Fermi energy of the
FM electrodes and the lowest unoccupied levels of the GF.
\begin{figure}
\centerline{\includegraphics[width=0.95\linewidth]{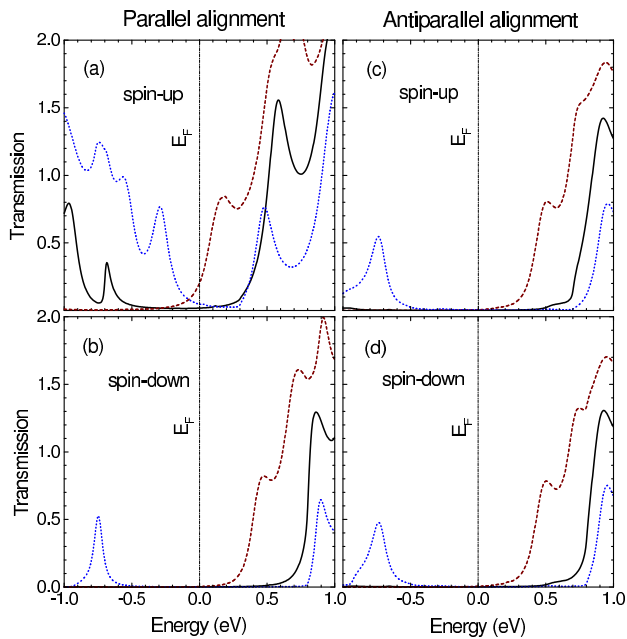}}
\caption{Transmission coefficients at $V_a$=0.1 V for FM/GF/FM
junction with zigzag interfaces in the parallel and antiparallel
magnetized states. Dashed, solid, and dotted curves correspond to
$V_G=-0.4$, 0.0 and 0.4 V, respectively.}
\end{figure}

\begin{figure}
\centerline{\includegraphics[width=0.95\linewidth]{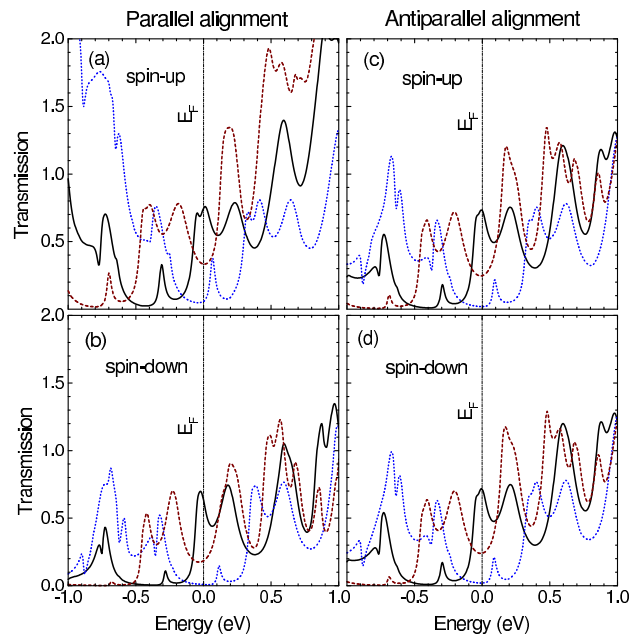}}
\caption{Transmission coefficients at $V_a$=0.1 V for FM/GF/FM
junction with armchair interfaces in the parallel and antiparallel
magnetized states. Dashed, solid, and dotted curves correspond to
$V_G=-0.4$, 0.0 and 0.4 V, respectively.}
\end{figure}

Interestingly, as shown in Fig. 3, the calculated transmission
coefficients of two spin channels in both magnetic alignments are
significantly larger than the transmission spectra of Fig. 2 near
the Fermi energy and thus according to Figs. 5(a)-(c) our theory
predicts a linear behavior in the $I$-$V$ characteristics at low
bias voltages for the junction with armchair interfaces. In
contrast to the spin currents of the junction with zigzag
interfaces, none of the spin currents exhibits a gap in the
$I$-$V$ characteristics of Fig. 5, which is due to the existence
of the resonant states near the Fermi energy in the junction with
armchair interfaces.

The main factor of difference in the calculated transmission
spectra arises due to the quantum interference effects. In the
case of armchair interfaces, the current flow mechanism
corresponds to the resonant tunneling which can be attributed to
Fabry-P\'{e}rot interference of electronic states partially
reflected from the electrodes, due to the finite length of the
graphene sample \cite{Blanter}. In the case of zigzag interface,
however, the interference between electronic waves scattered by
the carbon atoms along the FM/GF interfaces becomes important,
some resonances might completely disappear, and the transmission
curve changes (compare Fig. 2 with Fig. 3). Therefore, the current
flow mechanism is tunneling. The physical meaning of the
interference effect is that the electron waves in the GF which
come from the FM electrode may suffer a phase shift. Thus, a
constructive or destructive interference in the propagation
process of the electron through the GF can occur.

Another interesting feature of the junctions is the
magnetoresistance ratio which can be obtained from the $I$-$V$
curves using the definition: $\mathrm{MR}\equiv(I_p-I_a)/I_p$,
where $I_{p,a}$ are the total currents in the parallel and
antiparallel alignments of magnetizations in the FM electrodes,
respectively. In Figs. 4(d) and 5(d) we have shown the MR ratios
for the two junction at positive and negative gate voltages. It is
clear that the MR in the junction with zigzag interfaces has high
values in comparison with the other junction. The origin of
magnetoresistance effect is the difference in the total currents
which is related to the asymmetry of surface density of states of
the FM electrodes for spin-up and spin-down electrons and the
quantum tunneling phenomenon through the GF. In the parallel
alignment, minority (majority) electrons in the left FM electrode
go into the minority (majority) states of the right one by
tunneling through the GF. If, however, the two FM electrodes are
magnetized in the opposite directions, the minority (majority)
electrons from the left electrode seek empty majority (minority)
states in the right electrode. Consequently, the parallel
arrangement gives much higher total current through the GF than
does the antiparallel arrangement in the selected voltage interval
except for zero gate voltage at high (low) applied voltages for
the junction with zigzag (armchair) interfaces.

We also studied the effect of $t_{LC}$ on the spin currents and MR
ratio. The results showed that in both junctions, the values of
currents increase with increasing $t_{LC}$ (from 0.5 to 1 eV),
while the MR values do not change considerably due to its relation
to relative change of total currents between both magnetic
alignments.

\begin{figure}
\centerline{\includegraphics[width=0.95\linewidth]{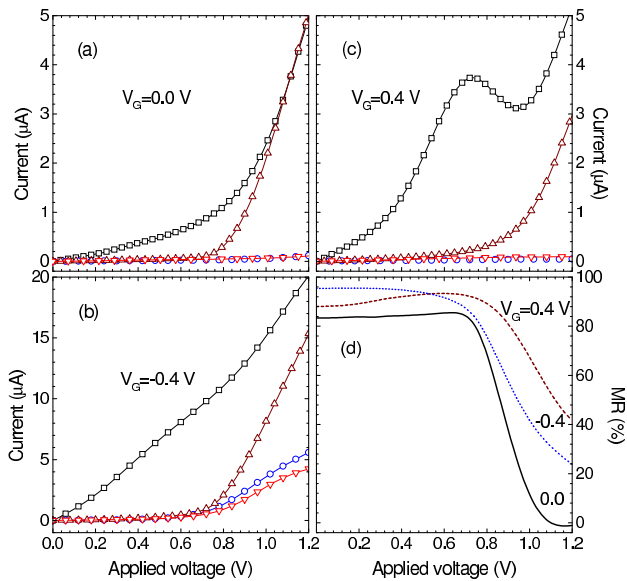}}
\caption{(a)-(c) Spin currents as a function of applied voltage
for FM/GF/FM junction with zigzag interfaces. Square and circle
(up-triangle and down-triangle) curves correspond to the spin-up
and spin-down currents in the parallel (antiparallel) alignment.
(d) MR as a function of applied voltage at different gate
voltages.}
\end{figure}

\begin{figure}
\centerline{\includegraphics[width=0.95\linewidth]{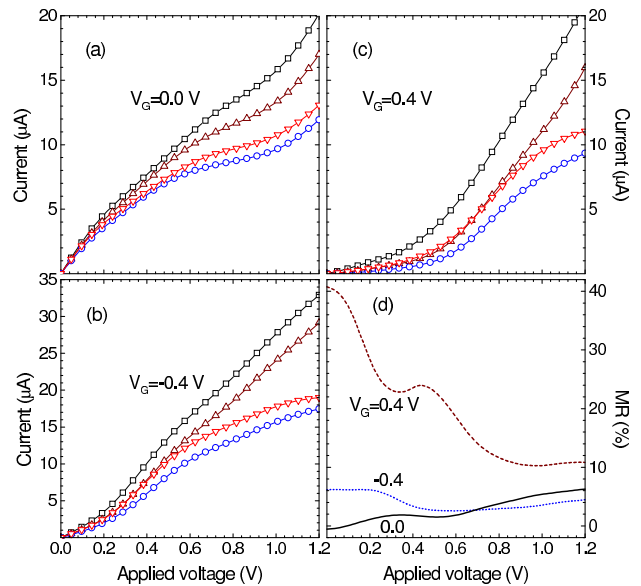}}
\caption{(a)-(c) Spin currents as a function of applied voltage
for FM/GF/FM junction with armchair interfaces. Square and circle
(up-triangle and down-triangle) curves correspond to the spin-up
and spin-down currents in the parallel (antiparallel) alignment.
(d) MR as a function of applied voltage at different gate
voltages.}
\end{figure}

Now we present a discussion on the spin states at the edges of GF.
As we pointed out, zigzag edge GNRs present half-metallic behavior
when homogeneous electric fields are applied across the ribbons
(in our model, along the $y$ direction of Fig. 1(a)). This
behavior is related to the existence of electronic states at the
Fermi energy spatially localized at the zigzag edges. The
ground-state spin configuration at zero electric field corresponds
to an opposite spin orientation across the ribbon between the
spin-polarized edge states. That is, the spins on the outermost
sites of each zigzag edge are parallel, resulting in one edge
polarized with spin up and the other polarized with spin down, and
the total zigzag-GNR spin is equal to zero \cite{Son}. In
contrast, the armchair edge GNRs do not show such spin states due
to the delocalized nature of the frontier orbitals \cite{Zhou}.
Theoretical studies have revealed that in zigzag GNRs; (i) the
critical electric field to achieve half-metallicity decreases with
the increase of ribbon width, (ii) the ribbon remains
half-metallic only at a limited range of electric field, and (iii)
the half-metallicity will be destroyed by a too strong electric
field and the ribbon becomes nonmagnetic \cite{Son,Kan,Hod}.

From a theoretical point of view, in order to observe the spin
states due to the edge-localized states in the zigzag GNRs, one
can use of first-principles calculations, which can predict the
electrical and magnetic properties of a material from the atomic
number and mass of its constituent atoms
\cite{Son,Kim,Kan,Hod,Pisani}, or include the electron-electron
interaction in the total Hamiltonian \cite{Guo,Rossier}. The
structure used in this study is analogous to that of Ref. [16].
However, we have used of the single-band tight-binding Hamiltonian
in the absence of interactions, because in this work we are
interested in the effects of the FM/GF interfaces on the spin
transport. Hence, the results do not yield the spin states at the
edges of zigzag ribbon [Fig. 1(a)]. According to the above
discussion, such a restriction does not invalidate our model for
the zigzag edge junction, since the electronic states
corresponding to the edge states, as it is clear from Fig. 3, are
far from the Fermi energy. The present theory demonstrates well
the effect of magnetic states of the electrodes on the electron
conduction through the GF, and also reveals well the role of
contact geometry in the spin-polarized transport through the two
above junctions. We should note that, if we include the
interactions in our Hamiltonian, the magnetic states corresponding
to the edges may affect the spin currents and the MR.

\section{Conclusion}

In this work, spin polarized transport of armchair and zigzag
interfaces between a GF and FM electrodes (Fig. 1) is investigated
in the tight-binding approximation. The results indicate that the
atomic arrangement at the interface between the FM electrode and
the GF has a dominant effect on the electron conduction through
the graphene-based magnetic junctions. The planar FM/GF/FM
junction with zigzag interfaces exhibits a spin-valve effect with
MR values as high as 95\% and negative differential resistance
features for a single spin channel. In the case of armchair
interface, both spin channels are in on state with low MR ratios.
In addition, the results of bias and gate voltage dependence of
the spin currents and MR have been reported.

In this study, we ignored the effects of spin states at the edges
of GF and the magnetic anisotropy of the FM electrodes due to the
reduced dimensionality and the geometry of the system. Although,
these factors can affect the spin-dependent transport, the present
study advances the fundamental understanding of interfacial
effects in the graphene-based magnetic junctions and suggests that
the junction with zigzag interfaces is an interesting candidate
for application in the magnetic memory cells and spintronic
devices.

\end{document}